\documentclass[a4paper,11pt]{article}
\usepackage{latexsym}	
\usepackage{amsmath}
\usepackage{amsthm}

\def\ba{\begin{eqnarray}}
\def\ea{\end{eqnarray}}

\def\ba{\begin{eqnarray}}
\def\ea{\end{eqnarray}}

\def\lb{\label}
\def\be{\begin{equation}}
\def\ee{\end{equation}}

\theoremstyle{plain}

\begin{document}

\title{Classical fermion dynamics in presence of non abelian monopoles}
\author{Alejandro Morano \thanks{Departamento de F\'isica, FCEyN, Universidad de Buenos Aires, Buenos Aires, Argentina
a.g.morano@gmail.com and osantil@dm.uba.ar.} and Osvaldo P. Santill\'an \thanks{IMAS, FCEyN, Universidad de Buenos Aires, Buenos Aires, Argentina
firenzecita@hotmail.com and osantil@dm.uba.ar.}}
\date {}
\maketitle
\begin{abstract}
In the present letter, the dynamics of a spin one-half particle with non abelian charge, interacting with a non abelian monopole like configuration, is studied.
In the non spinning case, these equations correspond to the Wong ones \cite{wong}, and the associated dynamics has been extensively investigated in \cite{ruso1}-\cite{ruso2}. 
The classical limit of a spinning particle in an abelian was considered in \cite{morano1}, where an interpretation of the celebrated Callan-Rubakov effect \cite{callanrubakov}-\cite{rubakov2} was obtained in purely classical terms.
The present work studies an interpolating situation, in which spin and non abelian charges are turned on. The corresponding equations are obtaining by taking into account some
earlier works about spinning particles \cite{vanholten}-\cite{moranoviejo2}. The conservation laws are studied in this context, and it is found that energy and some generalization of angular momentum
are conserved just in particular limits. This is a reflection that the non abelian interaction contains gauge field quartic interaction, which spoils the symmetries leading to these conservation laws. The precession of the particle is analyzed and compared with its abelian counterpart in this limit.

\end{abstract}
\section{Introduction}
Non abelian gauge theories are one of the most important subjects in modern theoretical physics, as they may explain phenomena such quark confinement
and several aspects of hadron physics. One of the aspects that makes them harder to understand is that quarks seemingly are not part of the asymptotic states of the theory.
It was suggested at some point that non abelian monopoles \cite{monopolos1}-\cite{monopolos2}  may play an important role for confinement, since they may help for realizing a dual Meissner effect in which the quarks become the analog of confined magnets. In this scenario, the monopoles are decoupled at low energy due to their large mass \cite{thooft}-\cite{mandelstam}. As far as the authors know, this idea has not yet successfully implemented. However, the study of flux lines, confinement and dualities has been proved to be fruitful in the supersymmetric context. Supersymmetry has not been discovered yet, nevertheless, these studies are an interesting 
theoretical laboratory for understanding the possible phases that a gauge theory may realize \cite{sw1}-\cite{seiberg4}.

The present work is focused on the classical dynamics of a particle with one-half spin in presence of non abelian fields, in particular monopole configurations. We underline that either the abelian limit or the spinless case have been considered in the precedent investigations. In particular, the references  \cite{ruso1}-\cite{ruso2} have found several classical solutions for spinless particles in presence of non abelian monopoles and the reference \cite{morano1} provides a deduction of the famous Callan-Rubakov result in purely classical terms. At first sight, one may wonder about the possible applications of such studies, as quarks are not free and the quantum effects are believed to play a significant role in the dynamics.
However, some motivations may be given. An example is the study of a non abelian plasma \cite{litim}-\cite{litim12} at very high densities or temperatures.  These studies may have applications in the very early universe, the dynamics of neutron stars or supernova explosions. In addition, the study of a electroweak plasma may have interesting applications for modelling baryon number violation at these cosmological stages.
Several topics about the dynamics of monopoles are covered in  \cite{shifmanbook1}-\cite{shnir}.  

The present work is organized as follows. In section 2 the equations of motion are derived, and the abelian limit and the spinless case 
are also discussed. In section 3 and 4 the energy and generalized angular momentum are discussed, and it is shown that these quantities
are conserved only in specific situations. An example is presented in section 5, where the spin and isotopic charge precession are studied in
some detail. This section also contains a discussion of the presented results.

\section{Derivation of the equations of motion}

\subsection{An abelian probe with spin in a monopole field}

Before deriving the equations of motion of a spinning non abelian particle in the field of a non abelian monopole, it  may be advisable to study some simpler situations.
In an abelian electromagnetic field, and in the non relativistic limit, the equations of motion of a charged particle are given by
\be\lb{mm1}
m\ddot{\overline{r}}=Qe(\overline{E}+\overline{v}\times \overline{B})+\frac{Qe \widetilde{g}}{2m} \nabla (\overline{S}\cdot \overline{B}),
\ee
\be\lb{mm2}
\frac{d\overline{S}}{dt}=\frac{Q e\widetilde{g}}{2m} (\overline{S}\times \overline{B}),
\ee
where $Qe$ is the particle charge, $m$ its mass and $\widetilde{g}$ is the g-factor giving the connection between the 
spin of the particle and the spin dependent magnetic moment. The back reaction of the particle on the source is neglected, which means that the gauge fields  are not affected by the dynamics. The second term of (\ref{mm1}) represents the interaction of the spin and the magnetic field
and the equation (\ref{mm2}) represents Larmor spin precession. These equations are well known. It is also known that the energy 
\be\lb{enershi}
E=\frac{m}{2}[\dot{r}^2+r^2(\dot{\theta}^2+\sin^2\theta \dot{\phi}^2)]-\frac{Qe\widetilde{g}}{2m}\overline{B}\cdot \overline{S},
\ee
is conserved for a magnetic field independent on time. Such is the case for the abelian monopole field
\be\lb{am}
\overline{B}=\frac{g \overline{r}}{r^3},
\ee
with $g$ the monopole magnetic charge, which satisfies the Dirac quantization $2eg=n$. 

The issues related to angular momentum conservation are more tricky. Even taking into account the rotational symmetry of the monopole magnetic field,
the angular momentum is not conserved as the spin-magnetic field coupling spoils this symmetry. However, the following quantity 
\be\lb{jota}
\overline{J}=mr^2[\dot{\theta}\hat{\phi}-\dot{\phi}\sin\theta \hat{\theta}]+\overline{S}-\frac{Qn}{2} \hat{r}.
\ee
has dimensions of an angular momentum and is conserved. Here the standard spherical coordinates were employed, in which the axis $\hat{x}$ corresponds to $2\theta=\pi$ and $\phi=0$.
If a particle incident to the monopole is considered, with impact parameter $d$ and velocity $v$ in the negative $\hat{x}$ direction, then the value of $\overline{J}$ is specified as
\be\lb{initio}
\overline{J}=mdv\widehat{z}-\frac{Qn}{2} \widehat{x}+\overline{S}_0,
\ee
with $\overline{S}_0$ the initial spin direction. On the other hand, (\ref{jota}) implies that
$$
m^2r^4(\dot{\theta}^2+\sin^2\theta \dot{\phi}^2)=j^2+s^2-2 \overline{J}\cdot\overline{S}-\frac{Q^2n^2}{4},
$$
where $j$ and $s$ indicates the modulus of their capital letter counterparts. By use of the value (\ref{initio}) it follows that
$$
m^2r^4(\dot{\theta}^2+\sin^2\theta \dot{\phi}^2)=m^2 d^2 v^2-2\overline{S}\cdot[\overline{S}_0-\overline{S}]+2mdv [(S_0)_z-S_z]-Qn [(S_0)_x-S_x]
$$
By eliminating, by use of the last expression, the angular part in the energy $E=mv^2/2$, it is arrived to
\be\lb{enershi2}
\dot{r}=\pm \sqrt{v^2-\frac{2m^2 d^2 v^2+2(2mdv \widehat{z}-Qn \widehat{x}-2\overline{S})\cdot(\overline{S}_0-\overline{S})-Q n \widetilde{g}\overline{S}\cdot \hat{r}}{2m^2 r^2}},
\ee
The spin equations are
$$
\frac{dS_x}{dt}=\frac{Q n \widetilde{g}}{m r^2}(S_y \cos\theta-S_z \sin \theta \sin \phi),
$$
$$
\frac{dS_y}{dt}=\frac{Q n \widetilde{g}}{m r^2}(S_z\sin\theta\cos\phi-S_x\cos\theta),
$$
\be\lb{crespin}
\frac{dS_z}{dt}=\frac{Q n \widetilde{g}}{m r^2}(S_x\sin\theta\sin\phi-S_y\sin\theta\cos\phi).
\ee
The angular momentum in spherical coordinates reads
\be\lb{jota2}
\overline{J}=mr^2[\dot{\theta}\widehat{\phi}-\dot{\phi}\sin\theta \widehat{\theta}]+\overline{S}-\frac{Qn}{2} \widehat{r},
\ee
From here and the initial value, it is found that
\be\lb{fipunto}
\frac{d\phi}{dt}=\frac{2mdv+Qn\cos\theta+2S_{0z}-2S_z}{2mr^2\sin^2\theta},
\ee
\be\lb{titapunto}
\frac{d\theta}{dt}=\frac{Qn\sin\theta\sin\phi+2S_{0y}-2S_y}{2mr^2\cos\phi}+\cot\theta \tan \phi \bigg[\frac{mdv+Qn\cos\theta+2S_{0z}-2S_z}{2mr^2}\bigg]
\ee
From (\ref{enershi2}) it is seen that there is a spin dependent radius of return defined by 
$$
r^2=\frac{2m^2 d^2 v^2+2(2mdv \widehat{z}-Qn \widehat{x}-2\overline{S})\cdot(\overline{S}_0-\overline{S})-Q n\widetilde{g} \overline{S}\cdot \hat{r}}{2m^2 v^2}.
$$
The intention is to estimate the minimum value of this quantity, that is, the closest radius the particle approaches. The precise value of $\hat{r}$
for the particle can be found only by solving the dynamics. For the estimation, one may assume that $\overline{S}\cdot \hat{r}=s$, that is, that they are parallel.
This gives the minimal contribution of the last term. On the other hand, the extremal values of the remaining quantity of the right hand side with respect to $\overline{S}$ leads to
$$
r_e^2\sim\frac{2m^2 d^2 v^2+(2mdv \widehat{z}-Qn\widehat{x})\cdot\overline{S}_0-Q n\widetilde{g} s}{2m^2 v^2}.
$$
The term $(mdv \widehat{z}-Qeg \widehat{x})\cdot\overline{S}_0$ takes the minimun when the vectors are anti parallel, the result is
$$
r_e^2\sim \frac{2m^2 d^2 v^2-s\sqrt{4m^2d^2v^2+Q^2n^2}-Qn\widetilde{g}}{2m^2 v^2}.
$$
The particles will be expelled from the center when the quantity given above is positive, and this implies that the impact parameter should be
$$
d> d_c=\sqrt{\frac{s-Q^2n^2+\sqrt{s^2+Qn\widetilde{g}-Q^2 n^2}}{4m^2 v^2}}
$$
The equality defines an approximate critical value $d_c$ for which the particles will not be deflected.
This is of course a rough approximation, as there some uncertainty about the value of $\hat{r}$ involved, which is reflected that 
for large $Qn$ the quantity in the square root may be negative. However, for moderate values
the  corresponding scattering cross section goes roughly as
$$
\sigma\sim \pi d_c^2\sim \frac{\pi}{2}\bigg(\frac{1}{mv}\bigg)^2(s-Q^2n^2+\sqrt{s^2+Qn\widetilde{g}-Q^2 n^2}),
$$
which, for $3v\sim 1$ and a mass similar to some typical fermion of the standard model such as an electron, is of the order of the Callan-Rubakov estimation \cite{callanrubakov}-\cite{rubakov2}. These classical considerations were employed in \cite{morano1}, although with slightly different methodology.

   Another important fact about spinless monopole scattering is the existence of rainbow and glory effects, as reviewed for instance in \cite{rainbow1}. For an spinless particle, the conserved quantity (\ref{jota}) becomes
  \be\lb{jotas}
\overline{J}=mr^2[\dot{\theta}\hat{\phi}-\dot{\phi}\sin\theta \hat{\theta}]-\frac{Qn}{2} \hat{r}.
\ee 
From here it is seen that
\be\lb{angcons}
2\overline{J}\cdot \hat{r}=Qn.
\ee
This means that the trajectory of the particle is inside a cone with a tip angle
$$
\cos\alpha=\frac{\overline{J}\cdot \hat{r}}{|\overline{J}|}.
$$
Therefore if some new polar coordinates $(r, \alpha, \beta)$ are chosen such that $\overline{J}$ is taken as the polar axis, then $\dot{\alpha}=0$
and the problem can be reduced to an effective central one \cite{rainbow1}. The use of this description allows for determining those effects.

When spin is turned on, this picture is more complicated. In this case
\be\lb{angcons}
2(\overline{J}-\overline{S})\cdot \hat{r}=Qn.
\ee
In this case the spin is precessing, thus the last equation does not implies that the particle moves inside a cone.
Thus, there is a third angle involved in the problem and the reduction to a central effective problem, if possible, is not
easy to be found.

\subsection{A non abelian spinless charge in a non abelian field}
On the other hand, the relativistic dynamics of an spinless particle in a non abelian gauge field is described by the Wong equations \cite{wong}
\be\lb{w1}
m\dot{u}_\mu=g F^a_{\mu\nu} u^{\nu} I^a,
\ee
\be\lb{w2}
\frac{d I^a}{dt}=g f^{abc}I^b[A_0^c+\overline{v}\cdot \overline{A}^c].
\ee
Here the signature convention  $(1,-1,-1, -1)$ is employed and $u^\mu$ is the relativistic four velocity $u^\mu=dx^\mu/d\tau$, with $\tau$ the proper time. The non abelian charge is characterized by the vector $I^a$, which may depend on $\tau$ during the particle evolution. The external non abelian gauge field is denoted by $A_\mu$, and the  non abelian stress tensor is calculated by means of the formula
$$
F_{\mu\nu}=\partial_{\mu} A_{\nu}-\partial_{\nu} A_{\mu}-ig [A_\mu, A_{\nu}].
$$
By employing generators $T^a$ of the gauge group such that $[T^a, T^b]=i f_{abc}T^c$ with $f_{abc}$ real constants, the stress tensor may be expressed as
$$
 F_{\mu\nu}=F_{\mu\nu}^a T^a, \qquad  F^a_{\mu\nu}=\partial_{\mu} A^a_{\nu}-\partial_{\nu} A^a_{\mu}+g f_{abc} A^b_\mu A^c_{\nu}.
$$
In particular, in the the $SU(2)$ case considered below $a=1,2,3$,  $f_{abc}=\epsilon_{abc}$ and $2T^a=\sigma^a$,  with $\sigma^a$ the standard Pauli matrices.
In general, the covariant derivative of given non abelian field $\varphi$ is expressed as
$$
D_\mu\varphi=\partial_\mu\varphi-ig [A_\mu, \varphi].
$$
In these terms, the equations of motion (\ref{w1})-(\ref{w2}) can be shown to be equivalent to
\be\lb{ww1}
\frac{d}{dt}\bigg(\frac{m\overline{v}}{\sqrt{1-v^2}}\bigg)=g(\overline{E}^a+\overline{v}\times \overline{B}^a)I^a,
\ee
\be\lb{ww2}
\frac{d}{dt}\bigg(\frac{m}{\sqrt{1-v^2}}\bigg)=g\overline{v}\cdot\overline{E}^aI^a,
\ee
\be\lb{ww3}
\frac{d I^a}{dt}=g f^{abc}I^b[A_0^c+\overline{v}\cdot \overline{A}^c].
\ee
 Here the external non abelian electric and magnetic fields $\overline{E}^a$
and $\overline{B}^a$ are given by
\be\lb{elec1}
E^a_i=F_{0i}^a=\partial_{0} A^a_{i}-\partial_{i} A^a_{0}+g f_{abc} A^b_0 A^c_{i},\qquad B_i^a=\frac{\epsilon_{ijk}}{2} F^a_{jk}.
\ee
The interpretation of these equations is clear. The first are the relativistic Newton equations in a non abelian gauge field. The second is the kinetic energy variation, as a consequence of the work of the electric force on the particle. The last ones represent the evolution of the charge vector $I^a$ due to the particle dynamics. Classical solutions of this equations were
studied, for instance, in \cite{ruso1}-\cite{ruso2}.

\subsection{A non abelian charge with spin in a non abelian field}
The equations of motion for an spinning probe particle in a general electromagnetic field will be taken, following the references \cite{vanholten}-\cite{moranoviejo2}, as follows
\be\lb{m1}
\frac{d}{dt}\bigg(\frac{m\overline{v}}{\sqrt{1-v^2}}\bigg)=g(\overline{E}^a+\overline{v}\times \overline{B}^a)I^a+\frac{\widetilde{g} g}{2m} I^a S_l[\nabla B^a_l-g f^{abc}B^b_l\overline{A}^c],
\ee
\be\lb{m2}
\frac{d\overline{S}}{dt}=\frac{\widetilde{g} g}{2m}I^a (\overline{S}\times \overline{B}^a),
\ee
\be\lb{m3}
\frac{d I^a}{dt}=g f^{abc}I^b[A_0^c+\overline{v}\cdot \overline{A}^c]+\frac{\widetilde{g} g}{2m} \epsilon^{abc} I^b \overline{S} \cdot \overline{B}^c.
\ee
The last system interpolate between the two situations described above namely, the abelian case and the non abelian Wong equations. Here the quantity $\widetilde{g}$
generalizes the g-factor for the abelian case, and no particular assumption is made about its value, except that it does not take very large values. The abelian limit follows by choosing the structure
constants $f^{abc}=0$ and the second by putting $\overline{S}=0$ in the last equations. The only term that requires some comment, as it is not a type of term
discussed in the previous section, is the last one in \eqref{m3}. The proportionality coefficient
is chosen as $\frac{\widetilde{g} g}{2m}$ because, if the effect of the gauge field $A_\mu^a$ is neglected locally, the resulting equations reduce to the formulas (4.7)
of reference \cite{moranoviejo2}.

 The next issue is to understand in which situations there are conserved quantities for the given system of equations \eqref{m1}-\eqref{m3}. The non abelianity of the fields may alter the intuition about the conserved quantities, as there are quartic gauge field interactions which are absent in the abelian case.

\section{The energy balance equation}
By multiplying (\ref{m1}) by $v=\dot{\overline{r}}$, the following expression for the time variation of the kinetic energy of the particle is obtained
\be\lb{m4}
\frac{d}{dt}\bigg(\frac{m}{\sqrt{1-v^2}}\bigg)=g\overline{v}\cdot\overline{E}^aI^a+\frac{\widetilde{g} g}{2m} I^a S_l[\overline{v}\cdot \nabla B^a_l-g f^{abc}B^b_l(\overline{v}\cdot\overline{A}^c)].
\ee
On the other hand, the generalization of the potential energy in an electric field for the non abelian case will be $I^a A_0^a$. Its variation is 
$$
\frac{d(I^a A_0^a)}{dt}=\frac{dI^a}{dt} A_0^a+I^a[\partial_0 A_0^a+\overline{v}\cdot \nabla \overline{A}_0^a].
$$
By using (\ref{elec1}) in order to eliminate the term $\nabla \overline{A}_0^a$ and also by taking into account (\ref{m3}) it is found that
$$
\frac{d(I^a A_0^a)}{dt}=g f^{abc}A_0^aI^b[A_0^c+\overline{v}\cdot \overline{A}^c]+\frac{\widetilde{g} g}{2m} \epsilon^{abc} I^b \overline{S} \cdot \overline{B}^cA_0^a
$$
$$
+I^a[\partial_0 A_0^a+\overline{v}\cdot (\partial_0 \overline{A}^a-\overline{E}^a+g f_{abc} A_0^b \overline{A}^c)].
$$
In the last expression, the first term is zero due to the anti-symmetry of $f_{abc}$ with respect of its indices. The second term cancels the last one. Thus  
$$
\frac{d(I^a A_0^a)}{dt}=\frac{\widetilde{g} g}{2m} \epsilon^{abc} I^b \overline{S} \cdot \overline{B}^cA_0^a
+I^a[\partial_0 A_0^a+\overline{v}\cdot (\partial_0 \overline{A}^a-\overline{E}^a)].
$$
The sum of the last expression multiplied with $g$, together with (\ref{m4}), allows to conclude that
$$
\frac{d}{dt}\bigg(\frac{m}{\sqrt{1-v^2}}+gI^a A_0^a\bigg)=\frac{\widetilde{g} g}{2m} I^a S_l[\overline{v}\cdot \nabla B^a_l-g f^{abc}B^b_l(\overline{v}\cdot\overline{A}^c)]
$$
$$
+\frac{\widetilde{g} g^2}{2m} \epsilon^{abc} I^b \overline{S} \cdot \overline{B}^cA_0^a
+g I^a[\partial_0 A_0^a+\overline{v}\cdot (\partial_0 \overline{A}^a)].
$$
Note that the terms proportional to the non abelian electric field $E^a_i$ cancelled each other. The last formula may be expressed as
$$
\frac{d}{dt}\bigg(\frac{m}{\sqrt{1-v^2}}+gI^a A_0^a\bigg)=-\frac{\widetilde{g} g^2}{2m}f_{abc} I^aS_l B^b_l(\overline{v}\cdot\overline{A}^c)]+\frac{\widetilde{g} g^2}{2m} \epsilon_{abc} I^b \overline{S} \cdot \overline{B}^cA_0^a
$$
$$
+\frac{\widetilde{g} g}{2m} I^a S_l \frac{dB^a_l}{dt}+g I^a[\partial_0 A_0^a+\overline{v}\cdot (\partial_0 \overline{A}^a)].
$$
On the other hand, (\ref{m2}) shows that $\dot{\overline{S}}\cdot \overline{B}=0$ and therefore
$$
\frac{d}{dt}\bigg(\frac{m}{\sqrt{1-v^2}}+gI^a A_0^a-\frac{\widetilde{g} g}{2m} I^a B^a_l S_l \bigg)=-\frac{\widetilde{g} g^2}{2m}f_{abc} I^aS_l B^b_l(\overline{v}\cdot\overline{A}^c)]+\frac{\widetilde{g} g^2}{2m} \epsilon_{abc} I^b \overline{S} \cdot \overline{B}^cA_0^a
$$
$$
-\frac{\widetilde{g} g}{2m} B^a_l S_l \frac{dI^a}{dt}+g I^a[\partial_0 A_0^a+\overline{v}\cdot (\partial_0 \overline{A}^a)].
$$
By further taking into account (\ref{m3}) and the anti-symmetry of $f_{abc}$ the last expression becomes
$$
\frac{d}{dt}\bigg(\frac{m}{\sqrt{1-v^2}}+gI^a A_0^a-\frac{\widetilde{g} g}{2m} I^a B^a_l S_l \bigg)=-\frac{\widetilde{g} g^2}{2m}f_{abc} I^aS_l B^b_l(\overline{v}\cdot\overline{A}^c)]+\frac{\widetilde{g} g^2}{2m} \epsilon_{abc} I^b \overline{S} \cdot \overline{B}^cA_0^a
$$
$$
-\frac{\widetilde{g} g^2}{2m} f_{abc}B^a_l S_l I^b[A_0^c+\overline{v}\cdot \overline{A}^c]+g I^a[\partial_0 A_0^a+\overline{v}\cdot (\partial_0 \overline{A}^a)].
$$
The first and the fourth term cancel out, and the second and the third are proportional. Thus
$$
\frac{d}{dt}\bigg(\frac{m}{\sqrt{1-v^2}}+gI^a A_0^a-\frac{\widetilde{g} g}{2m} I^a B^a_l S_l \bigg)=\frac{\widetilde{g} g^2}{m} \epsilon_{abc} I^b \overline{S} \cdot \overline{B}^cA_0^a
$$
\be\lb{enbal}
+g I^a[\partial_0 A_0^a+\overline{v}\cdot (\partial_0 \overline{A}^a)].
\ee
This formula has important consequences. One may define the full energy of the particle as
\be\lb{etoto}
E=\frac{m}{\sqrt{1-v^2}}+gI^a A_0^a-\frac{\widetilde{g} g}{2m} \overline{S}\cdot \overline{B}^a I^a,
\ee
since it generalizes the abelian case. In fact, this is the sum of the kinetic energy, the generalization of the electric potential, and the Larmor one.
If there is a gauge for which $A^a_\mu$ is time independent, the last two terms cancel. However, this energy is not conserved due to the first term in (\ref{enbal}). This may be interpreted as an effect
due to the non linear interaction of the gauge field of the particle and the external one. 

The results given above show that energy is not conserved in general for non abelian probe particles moving in an abelian field. However, as it will be shown below, it is conserved for a monopole field in a $SU(2)$ gauge theory, in certain limit.

\section{The angular momentum balance equation}
The angular momentum is  not expected to be conserved either, even for a static field configuration. In the present section, the attention will be restricted to the class of static $SU(2)$ non abelian fields given by
\be\lb{gaugef}
A^a_0=\frac{f}{g r} n^a,\qquad A^a_i=\frac{(1-a)}{g r} \epsilon_{aij}n^j.
\ee
Here $a=a(r)$ and $f=f(r)$ are generic functions of the distance $r$ between the observation point and the non abelian source, whose explicit profile depends on the model of consideration. In the last expression, the unit vector
$$
n^a=\frac{x^a}{r},\qquad \dot{n}^a\cdot n^a=0,
$$
has been introduced, with $x^a$ a system of cartesian coordinates in $R^3$ describing the position of the probe particle. In the present case, the structure constants $f_{abc}=\epsilon_{abc}$. For the calculation to be performed below, it is convenient to write (\ref{m1})-(\ref{m3}) in terms
of (\ref{gaugef}), and the corresponding color electromagnetic fields. From (\ref{gaugef}) and (\ref{elec1}) it is found out that
\be\lb{electro}
E^a_i=-\frac{1}{g}\bigg[\bigg(\frac{f}{r}\bigg)' n^a n^i+\frac{af}{r^2}(\delta^{ai}-n^a n^i)\bigg],
\ee
\be\lb{magnetico}
B^a_i=-\frac{1}{g}\bigg[\frac{a^2-1}{r^2} n^a n^i+\frac{a'}{r}(\delta^{ai}-n^a n^i)\bigg].
\ee
By use of the last expressions, (\ref{gaugef}) and (\ref{m2}) it is found the following equation describing spin precession
\be\lb{espin}
\frac{d\overline{S}}{dt}=\frac{\widetilde{g}}{2m}\bigg[\frac{1-a^2}{r^2}(\overline{I}\cdot \overline{n})\overline{S}\times \overline{n}
-\frac{a'}{r}\overline{S}\times \overline{I}+\frac{a'}{r}(\overline{I}\cdot \overline{n})\overline{S}\times \overline{n}\bigg].
\ee
On the other hand, the formula (\ref{m3}) and the identity $\overline{v}=\dot{r}\overline{n}+r \dot{\overline{n}}$ imply that
$$
\frac{d \overline{I}}{dt}=\frac{f}{r}\overline{I}\times \overline{n}+(1-a)[(\overline{I}\cdot \overline{n})\dot{\overline{n}}-(\overline{I}\cdot \dot{\overline{n}})\overline{n}]
$$
\be\lb{isotopic}
+\frac{\widetilde{g}}{2m}  \bigg[\frac{1-a^2}{r^2}(\overline{S}\cdot \overline{n})\overline{I}\times \overline{n}-\frac{a'}{r}\overline{S}\times \overline{I}+\frac{a'}{r}(\overline{S}\cdot \overline{n})\overline{I}\times \overline{n}\bigg].
\ee
This, in particular, leads to the following formula
\be\lb{isotid}
\frac{d \overline{I}\cdot \overline{n}}{dt}=a(\overline{I}\cdot \dot{\overline{n}})
-\frac{\widetilde{g} }{2m} \frac{a'}{r}(\overline{S}\times \overline{I})\cdot \overline{n},
\ee
which will be employed below. 
Now, vector multiplication of (\ref{m1}) with respect to $\overline{v}$ leads to the following equation for the angular momentum of the particle
\be\lb{m5}
\frac{d\overline{L}}{dt}=g(\overline{r}\times \overline{E}^a+\overline{r}\times\overline{v}\times \overline{B}^a)I^a+\frac{\widetilde{g} g}{2m} I^a S_l[(\overline{r}\times \nabla) B^a_l-g f^{abc}B^b_l(\overline{r}\times\overline{A}^c)].
\ee
Here 
$$
\overline{L}=\frac{m\overline{r}\times \overline{v}}{\sqrt{1-v^2}},
$$
is the relativistic angular momentum of the particle.

There are several terms to be calculated in \eqref{m5}. First, from (\ref{electro})-(\ref{magnetico}) and by use of $\overline{v}=\dot{r}\overline{n}+r \dot{\overline{n}}$, it is found that
\be\lb{erste}
g(\overline{r}\times \overline{E}^a+\overline{r}\times\overline{v}\times \overline{B}^a)I^a=\frac{a f}{r}\overline{I}\times \overline{n}+(1-a^2)(\overline{I}\cdot \overline{n})\dot{\overline{n}}+\frac{da}{dt}(\overline{I}-(\overline{I}\cdot \overline{n})\overline{n}).
\ee
On the other hand, expression for the term $\nabla B_l^a$ is a bit cumbersome, but the expression for $(\overline{r}\times \nabla) B_l^a$ is significantly simpler. A direct calculation shows the validity of this identity for the present configuration
$$
\frac{\widetilde{g} g}{2m} I^a S_l(\overline{r}\times \nabla) B^a_l=\frac{\widetilde{g}}{2m} \bigg[\frac{1-a^2}{r^2}(\overline{S}\cdot \overline{n}) (\overline{n}\times \overline{I})
+\frac{1-a^2}{r^2}(\overline{I}\cdot \overline{n}) (\overline{n}\times \overline{S})
$$
$$
+\frac{a'}{r}(\overline{S}\cdot \overline{n}) (\overline{n}\times \overline{I})+\frac{a'}{r}(\overline{I}\cdot \overline{n}) (\overline{n}\times \overline{S})
\bigg].
$$
This expression describes the third term in (\ref{m5}). Comparison of the last terms of this formula with (\ref{espin}) and (\ref{isotopic}) allows to find a simpler expression for the third term, namely
\be\lb{zweite}
\frac{\widetilde{g} g}{2m} I^a S_l(\overline{r}\times \nabla) B^a_l=-\frac{d\overline{S}}{dt}-\frac{d\overline{I}}{dt}+\frac{f}{r}\overline{I}\times \overline{n}+(1-a)[(\overline{I}\cdot \overline{n})\dot{\overline{n}}+(\overline{I}\cdot \dot{\overline{n}})\overline{n}]-\frac{\widetilde{g}}{m}\frac{a'}{r} \overline{S}\times \overline{I}.
\ee
Finally, the last term in (\ref{m5}) can be calculated to give
$$
-\frac{\widetilde{g} g^2}{2m} I^a S_l \epsilon^{abc}B^b_l(\overline{r}\times\overline{A}^c)=\frac{\widetilde{g}}{2m}\frac{(1-a)a'}{r} \overline{I}\cdot (\overline{S}\times \overline{n})\overline{n}
$$
$$
+\frac{\widetilde{g}}{2m }(1-a)\bigg[\frac{1-a^2}{r^2}(\overline{S}\cdot \overline{n})\overline{I}\times \overline{n}-\frac{a'}{r}\overline{S}\times \overline{I}+\frac{a'}{r}(\overline{S}\cdot \overline{n})\overline{I}\times \overline{n}\bigg].
$$
The use of (\ref{espin}) and (\ref{isotopic}) again shows that the last formula is equivalent to the following one
$$
-\frac{\widetilde{g} g^2}{2m} I^a S_l \epsilon^{abc}B^b_l(\overline{r}\times\overline{A}^c)=\frac{\widetilde{g}}{2m}\frac{(1-a)a'}{r} \overline{I}\cdot (\overline{S}\times \overline{n})\overline{n}+(1-a)\frac{d \overline{I}}{dt}
$$
\be\lb{dritte}
-(1-a)\frac{f}{r}\overline{I}\times \overline{n}-(1-a)^2[(\overline{I}\cdot \overline{n})\dot{\overline{n}}-(\overline{I}\cdot \dot{\overline{n}})\overline{n}].
\ee
After these calculations, it is found that  sum of the formulas (\ref{erste})-(\ref{dritte}) imply that
$$
\frac{d\overline{L}}{dt}+\frac{d\overline{S}}{dt}=\frac{a f}{r}\overline{I}\times \overline{n}+(1-a^2)(\overline{I}\cdot \overline{n})\dot{\overline{n}}+\frac{da}{dt}(\overline{I}-(\overline{I}\cdot \overline{n})\overline{n})
$$
$$
-a\frac{d \overline{I}}{dt}+\frac{a f}{r}\overline{I}\times \overline{n}+a(1-a)[(\overline{I}\cdot \overline{n})\dot{\overline{n}}-(\overline{I}\cdot \dot{\overline{n}})\overline{n}]
+\frac{\widetilde{g}}{m }\frac{a'}{r} \overline{S}\times \overline{I}+\frac{\widetilde{g}}{2m}\frac{(1-a)a'}{r} \overline{I}\cdot (\overline{S}\times \overline{n})\overline{n}.
$$
It can not be concluded from this expression that the sum $\overline{J}=\overline{L}+\overline{S}$ is conserved. This is not peculiar however, as this quantity is not conserved in the abelian case either. Nevertheless, the use of (\ref{isotid}) allows to put the last expression in the following form
$$
\frac{d}{dt}(\overline{L}+\overline{S}-a\overline{I}-(1-a)(\overline{I}\cdot\overline{n})\overline{n})=\frac{2a f}{r}\overline{I}\times \overline{n}+(1-a^2)(\overline{I}\cdot \overline{n})\dot{\overline{n}}-(1-a)(\overline{I}\cdot \overline{n})\dot{\overline{n}}-(1-a)[a(\overline{I}\cdot \dot{\overline{n}})
-\frac{\widetilde{g} }{2m} \frac{a'}{r}(\overline{S}\times \overline{I})\cdot \overline{n}]\overline{n}
$$
$$
-2a\frac{d \overline{I}}{dt}+a(1-a)[(\overline{I}\cdot \overline{n})\dot{\overline{n}}-(\overline{I}\cdot \dot{\overline{n}})\overline{n}]
+\frac{\widetilde{g}}{m }\frac{a'}{r} \overline{S}\times \overline{I}+\frac{\widetilde{g}}{2m}\frac{(1-a)a'}{r} \overline{I}\cdot (\overline{S}\times \overline{n})\overline{n}.
$$
The use of (\ref{isotopic}) leads to
$$
\frac{d}{dt}(\overline{L}+\overline{S}-a\overline{I}-(1-a)(\overline{I}\cdot\overline{n})\overline{n})=(1-a^2)(\overline{I}\cdot \overline{n})\dot{\overline{n}}-(1-a)(\overline{I}\cdot \overline{n})\dot{\overline{n}}-(1-a)a(\overline{I}\cdot \dot{\overline{n}})\overline{n}
$$
$$
+a(1-a)[(\overline{I}\cdot \overline{n})\dot{\overline{n}}-(\overline{I}\cdot \dot{\overline{n}})\overline{n}]-2a(1-a)[(\overline{I}\cdot \overline{n})\dot{\overline{n}}-(\overline{I}\cdot \dot{\overline{n}})\overline{n}]
$$
$$
+\frac{\widetilde{g}}{m }\frac{a'}{r} \overline{S}\times \overline{I}+\frac{\widetilde{g}}{2m}\frac{(1-a)a'}{r} (\overline{I}\cdot (\overline{S}\times \overline{n}))\overline{n}+\frac{\widetilde{g} (1-a) }{2m} \frac{a'}{r}((\overline{S}\times \overline{I})\cdot \overline{n})\overline{n}
-\frac{\widetilde{g} a}{2m}  \bigg[\frac{1-a^2}{r^2}(\overline{S}\cdot \overline{n})\overline{I}\times \overline{n}
$$
$$
-\frac{a'}{r}\overline{S}\times \overline{I}+\frac{a'}{r}(\overline{S}\cdot \overline{n})\overline{I}\times \overline{n}\bigg].
$$
Several terms cancel each other out, and this formula simplify to 
$$
\frac{d}{dt}(\overline{L}+\overline{S}-a\overline{I}-(1-a)(\overline{I}\cdot\overline{n})\overline{n})=
\frac{\widetilde{g}}{m }\frac{a'(1-a)}{r} \overline{S}\times \overline{I}+\frac{\widetilde{g}}{2m}\frac{(1-a)a'}{r} ((\overline{S}\times \overline{n})\cdot \overline{I})\overline{n}
$$
$$
+\frac{\widetilde{g} (1-a) }{2m} \frac{a'}{r}((\overline{S}\times \overline{I})\cdot \overline{n})\overline{n}
-\frac{\widetilde{g} a}{2m}  \bigg[\frac{1-a^2}{r^2}(\overline{S}\cdot \overline{n})\overline{I}\times \overline{n}
+\frac{a'}{r}(\overline{S}\cdot \overline{n})\overline{I}\times \overline{n}\bigg].
$$
By use of algebra vector identities, this becomes 
$$
\frac{d}{dt}(\overline{L}+\overline{S}-a\overline{I}-(1-a)(\overline{I}\cdot\overline{n})\overline{n})=
\frac{\widetilde{g}}{m }\frac{a'(1-a)}{r} \overline{S}\times \overline{I}
-\frac{\widetilde{g} a}{2m}  \bigg[\frac{1-a^2}{r^2}
+\frac{a'}{r}\bigg](\overline{S}\cdot \overline{n})\overline{I}\times \overline{n}.
$$
This means that
\be\lb{obstruccion}
\frac{d\overline{M}}{dt}=\frac{\widetilde{g}}{m }\frac{a'(1-a)}{r} \overline{S}\times \overline{I}
-\frac{\widetilde{g} a}{2m}  \bigg[\frac{1-a^2}{r^2}
+\frac{a'}{r}\bigg](\overline{S}\cdot \overline{n})\overline{I}\times \overline{n},
\ee
where
\be\lb{eres}
\overline{M}=\overline{L}+\overline{S}-a\overline{I}-(1-a)(\overline{I}\cdot\overline{n})\overline{n}.
\ee
Therefore the quantity $\overline{M}$ is conserved in this case only if $a=0$ or $a=1$.
The case $a=0$ corresponds approximately to the asymptotic region, while the case $a=1$ describes the core of the
object. As the core is very small, the solution may be approximated with $a=0$ in almost all the region of the space.
Note that, in the abelian limit $\overline{I}\cdot\overline{n}=Qe$, the resulting conserved quantity coincides with
(\ref{jota}).

\section{Specific calculation}
In the following, the approximation $a=0$ will be employed, since describes approximately the field except for a very tiny region around the origin $r=0$, where the monopole is located. In this limit, the formula (\ref{isotid}) leads to 
\be\lb{constant}
\overline{I}\cdot \overline{n}=\overline{I}_0\cdot \overline{n}_0
\ee
where the $0$ represent the quantities at the beginning of the motion. The last formula implies that the projection
of the abelian charge on the direction joining the probe particle and the source is conserved. In addition, in this case, the energy balance equation implies that (\ref{enbal})
that
\be\lb{etoto2}
E=\frac{m}{\sqrt{1-v^2}}+gI^a A_0^a-\frac{\widetilde{g} g}{2m} \overline{S}\cdot \overline{B}^a I^a.
\ee
is conserved, as the term $\epsilon_{abc} I^b \overline{S} \cdot \overline{B}^cA_0^a$ is zero in this limit.
In fact, for $a=0$ the gauge fields are given by
$$
A^a_0=\frac{f}{g r} n^a,\qquad A^a_i=\frac{1}{g r} \epsilon_{aij}n^j.
$$
$$
E^a_i=-\frac{1}{g}\bigg(\frac{f}{r}\bigg)' n^a n^i,\qquad 
B^a_i=\frac{1}{g r^2} n^a n^i.
$$
These formulas imply that $\epsilon_{abc} I^b \overline{S} \cdot \overline{B}^cA_0^a$ is proportional to $\epsilon_{abc} n^b n^c$, which clearly vanishes.
In addition, it follows that
$$
\partial_j B_i^a=\frac{1}{g r^3} (\delta^{aj} n^i+\delta^{ai} n^j)-\frac{4n^i n^a n^j}{g r^3}.
$$
The equations of motion simplify to
$$
\frac{d}{dt}\bigg(\frac{m\overline{v}}{\sqrt{1-v^2}}\bigg)=-\bigg(\frac{f}{r}\bigg)'(\overline{I}\cdot\overline{n})\overline{n}+\frac{1}{ r^2} (\overline{I}\cdot\overline{n})(\overline{v}\times \overline{n})
+\frac{\widetilde{g}}{2mr^3} [(\overline{I}\cdot\overline{S})\overline{n}-3(\overline{I}\cdot\overline{n})(\overline{S}\cdot\overline{n})\overline{n}],
$$
$$
\frac{d\overline{S}}{dt}=\frac{\widetilde{g}}{2m r^2} (\overline{I}\cdot\overline{n}) (\overline{S}\times \overline{n}),
$$
$$
\frac{d \overline{I}}{dt}=\frac{f}{r}\overline{I}\times \overline{n}+\frac{\overline{I}\times \overline{v}\times \overline{n}}{r}+\frac{\widetilde{g}}{2m  r^2}  (\overline{S} \cdot \overline{n}) (\overline{I}\times\overline{n}).
$$
The conserved quantities are
$$
E=\frac{m}{\sqrt{1-v^2}}+\frac{ f}{ r} (\overline{I}\cdot\overline{n})-\frac{\widetilde{g}}{2m r^2} (\overline{S}\cdot \overline{n}) (\overline{I}\cdot\overline{n}).
$$
$$
\frac{m\overline{v}\times\overline{n}}{\sqrt{1-v^2}}=\frac{\overline{S}}{r}-\frac{\overline{M}}{r}+\frac{(\overline{I}\cdot\overline{n})\overline{n}}{r}.
$$
This implies that 
$$
(\overline{S}+\overline{I})\cdot \overline{n}=\overline{M}\cdot \overline{n}.
$$
Even taking into account (\ref{constant}), note that the last equation does not imply that $\overline{S}\cdot \overline{n}$ is conserved. 
The point is that $\overline{M}$ is a constant of motion, but its multiplication with $\overline{n}$ is not.
The conserved quantities may be written as
$$
E=\frac{m}{\sqrt{1-v^2}}+\bigg[\frac{f}{ r}  -\frac{\widetilde{g}}{2m r^2} (\overline{S}\cdot \overline{n})\bigg] (\overline{I}\cdot\overline{n}).
$$
$$
\frac{m\overline{v}\times\overline{n}}{\sqrt{1-v^2}}=\frac{\overline{S}}{r}-\frac{(\overline{S}\cdot\overline{n})\overline{n}}{r}-\frac{\overline{M}}{r}+\frac{(\overline{M}\cdot\overline{n})\overline{n}}{r}.
$$
Explicitly, for the non relativistic case, the energy is
$$
E=\frac{m}{2}(\dot{r}^2+r^2\dot{\theta}^2+r^2\sin^2\theta \dot{\phi}^2)+\bigg[\frac{ f}{r}
-\frac{\widetilde{g} }{2m r^2} (S_x\sin\theta\cos\phi+S_y\sin\theta\sin\phi+S_z\cos\theta)\bigg](\overline{I} \cdot\overline{n}).
$$
By taking this into account, the spin equations and the conservation of $M_i$ it is obtained the following linear system
$$
\frac{dS_x}{dt}=\frac{\widetilde{g} (\overline{I}\cdot\overline{n})}{2mr^2}(S_z\sin\theta\sin\phi-S_y\cos\theta),
$$
$$
\frac{dS_y}{dt}=\frac{\widetilde{g} (\overline{I}\cdot\overline{n})}{2m r^2}(S_x\cos\theta-S_z\sin\theta\cos\phi),
$$
$$
\frac{dS_z}{dt}=\frac{\widetilde{g} (\overline{I}\cdot\overline{n}) }{2m r^2} (S_y\sin\theta\cos\phi-S_x\sin\theta\sin\phi),
$$
\be\lb{sistemon}
\dot{\theta}=\frac{1}{m r^2}\bigg[(M_x-S_x)\sin\phi-(M_y-S_y)\cos\phi\bigg],
\ee
$$
\dot{\phi}=\frac{1}{mr^2\sin\theta}\bigg[(M_x-S_x)\cos\theta\cos\phi+(M_y-S_y)\cos\theta\sin\phi
-(M_z-S_z)\sin\theta\bigg],
$$

$$
\dot{r}=\pm\sqrt{\frac{2}{m}\bigg\{E-\frac{m}{2}(r^2\dot{\theta}^2+r^2\sin^2\theta \dot{\phi}^2)-\bigg[\frac{ f}{r}  
-\frac{\widetilde{g} }{2m r^2}(S_x\sin\theta\cos\phi+S_y\sin\theta\sin\phi+S_z\cos\theta)\bigg](\overline{I}\cdot\overline{n})\bigg\}} ,
$$
where the fourth and the fifth equations have to be replaced in the last one for eliminating $\dot{\theta}$ and $\dot{\phi}$. The last are a system of equation of first order for
the quantities describing the dynamics namely, $S_i$, $\theta$, $\phi$ and $r$. This is a closed system. It is clear that this system describes a particle which moves as an abelian particle
with charge $\overline{I}\cdot \overline{n}$, which is constant during the whole evolution. This does not implies however, that non abelian and abelian particles
can not be distinguished by the dynamics. This is discussed below.

\section{Discussion}
In the present work, some results that generalize certain aspects of fermion dynamics in presence of monopoles  such as the ones discussed in \cite{shnir}-\cite{weinberg} were presented. Some remarks are in order.
As pointed out in the book \cite{weinberg}, there is a symmetry for an abelian hedgehog monopole which involves a spatial rotation and non abelian charge rotation. Thus the action of $\overline{J}=\overline{L}+\overline{T}$
leaves invariant the solution. However, when a non zero spin is present, at classical level this is not a symmetry of the equations of motions, and instead the quantity (\ref{jota}), which is reproduced here by convenience
$$
\overline{J}=mr^2[\dot{\theta}\hat{\phi}-\dot{\phi}\sin\theta \hat{\theta}]+\overline{S}-\frac{Qn}{2} \hat{r},
$$
 is the one that is conserved \cite{morano1}. For non abelian charges, when the spin is taken $\overline{S}=0$ the conserved quantity is given by \cite{ruso1}
$$
\overline{M}_{s=0}=\overline{L}-a\overline{I}-(1-a)(\overline{I}\cdot\overline{n})\overline{n},
$$
which coincides with the quantity $\overline{M}$ defined in (\ref{eres}) in the case $\overline{S}=0$. In fact, this is consistent with (\ref{obstruccion}). This formula shows that the obstruction for $\overline{M}$
to be conserved is proportional to $\overline{S}$, thus conservation takes place for an spinless particle. 

The results of the present paper show that the quantity $\overline{M}$ when spin is present, strictly speaking, is not conserved for a non abelian monopole configuration. Neither does the energy (\ref{etoto}) namely
$$
E=\frac{m}{\sqrt{1-v^2}}+gI^a A_0^a-\frac{\widetilde{g} g}{2m} \overline{S}\cdot \overline{B}^a I^a.
$$
This non conservation may be interpreted as an effect of the non linear interaction of the field of the particle and the monopole one, which is absent for the abelian case.
This non conservations complicate the dynamics. However, as asymptotically
the profile $a(r)\to 1$ rapidly, and in this limit $\overline{M}$ is conserved, to postulate conservation of this quantity is approximately correct. In the inner region $a(r)\to 0$ and conservation also takes place.
Thus, if the profile is approximated by an step function $a(r)=\Theta(r-r_0)$ the problem can be divided in two regions $r\geq r_0$ and $r<r_0$ where $\overline{M}$ and $E$ are conserved.
The internal and external solutions have boundary conditions at the sphere $r=r_0$ which comes from the jump given in (\ref{obstruccion}). If the inner part region has small volume, then a simplifying 
assumption is to put $a\neq 0$ and employ conservation of these quantities. This leads to the system (\ref{sistemon}) in which, in addition, the quantity $\overline{I}\cdot \overline{n}$ is a constant of motion.

It is seem by comparing the obtained system (\ref{sistemon}) with the equations (\ref{mm1})-(\ref{titapunto}) that the dynamics of a spin particle in in the field of a non abelian monopole is equivalent, by interpreting the constant $\overline{I}\cdot\overline{n}$
as an effective charge, analogous to $2Qeg$ for the abelian case. However, the sign of the charge depends
on the initial conditions, and the force may be attractive or repulsive depending on the initial choice. In addition, there exist a regime $\overline{I}\cdot\overline{n}=0$ for which the charge 
does not experience the force of the monopole and just pass through with a straight line. This does not imply that the charges do not evolve in time. In fact,
in the limit $a=0$ the equations (\ref{isotopic}) are given by
\be\lb{isotopic3}
\frac{d \overline{I}}{dt}=\frac{f}{r}\overline{I}\times \overline{n}+[(\overline{I}\cdot \overline{n})\dot{\overline{n}}-(\overline{I}\cdot \dot{\overline{n}})\overline{n}]
+\frac{\widetilde{g}}{2m r^2} (\overline{S}\cdot \overline{n})\overline{I}\times \overline{n}.
\ee
It is seen that  even if $\overline{I}\cdot\overline{n}=0$ and $\overline{n}$ describes a straight line, this time derivative is not zero. Thus, there is charge precession even in this almost trivial case.

The fact that the charge is not invariant under reflections, that is, $\overline{I}\cdot\overline{n}\to- \overline{I}\cdot\overline{n}$ by a reflection by a plane with normal $\overline{n}$, has another possible consequence.
For a scattering beam aligned in the negative $\hat{x}$ direction, the value of $\overline{n}\sim- \hat{x}$, and this fixes the value of the charge of all the particles composing the beam as $-I_x$. The dynamics of such beam
is identical to one with abelian charge $Qge=-I_x$. On the other hand, if the beam is aligned in the positive $\hat{x}$ direction, and all the initial conditions are copied, then it works as a particle with charge opposite $Qeg=I_x$.
Therefore, if two beams coming from two opposite directions are incident over the monopole, with the same charge $\overline{I}$ and the same $S_{0y}$ and $S_{0z}$ and with opposite values of $S_{0x}$, the dynamics will not
be symmetric. This effect is different than in the abelian case, and distinguish the non abelian nature of the beam. 

An interesting topic may be the rainbow and glory effect, which is known to happen for the motion of spinless particles in the field of an abelian monopole \cite{rainbow1}.
This effect also is found in an opposite situation in which the monopole is not fixed, but it moves in the presence of a dyon \cite{rainbow2}-\cite{rainbow3}. As pointed out in (\ref{angcons}), if the spin is zero,
then the particle moves in the surface of a cone with angle dictated by the initial conditions. When spin is turned on, this is not the case, and the particle moves in a volume. This complicates considerably
the dynamics and the possibility of study these effects, as the previous simplification is of great help when dealing with these matters.

There is a possible additional difference between an abelian and non abelian dynamics, although the following argument is not completely rigorous. Consider a charge with spin in the middle of two identical abelian monopoles.
The formula (\ref{mm1}) shows that the force will be zero. However, if a direction dependent charge $\overline{I}\cdot\overline{n}$ is in the middle of two identical non abelian monopoles, there is no cancellation, as this seems
to act as a positive charge for one monopole and negative for the other. This argument is dubious however, as the conservation of $\overline{I}\cdot\overline{n}$ was obtained only for the limit $a=0$
in a single monopole configuration and may be violated for a two monopole field. In any case, we suggest that the dynamics of a non abelian charge in the field of two or more monopoles may 
be of academic interest. We leave this issue for a future publication.

\section*{Acknowledgments}
O. S is supported by CONICET, Argentina and by the Grant PICT 2020-02181.

\end{document}